\documentclass[a4paper,11pt,aps,nofootinbib]{revtex4}
\pdfoutput=1 
\usepackage{amsmath}
\usepackage{amssymb}
\usepackage{tikz}
\usetikzlibrary{positioning}
\usepackage{graphicx}
\usepackage{epstopdf}
\usepackage{multirow}
\usepackage{hyperref}
\usepackage{inputenc}
\newcommand{\X}{{\cal X}}
\begin{document}

\title{Classification of the Horndeski cosmologies via Noether Symmetries}

\author{Salvatore Capozziello}
\email{capozziello@na.infn.it}
\affiliation{Dipartimento di Fisica "E. Pancini", Universit\'a di Napoli
	\textquotedblleft{Federico II}\textquotedblright, Napoli, Italy,}
\affiliation{Gran Sasso Science Institute, Via F. Crispi 7, I-67100, L' Aquila,
	Italy,}
\affiliation{INFN Sez. di Napoli, Compl. Univ. di Monte S. Angelo, Edificio G, Via
	Cinthia, I-80126,
	Napoli, Italy.}

\author{Konstantinos F.	Dialektopoulos}
\email{dialektopoulos@na.infn.it}
\affiliation{Dipartimento di Fisica "E. Pancini", Universit\'a di Napoli
	\textquotedblleft{Federico II}\textquotedblright, Napoli, Italy,}
\affiliation{INFN Sez. di Napoli, Compl. Univ. di Monte S. Angelo, Edificio G, Via
	Cinthia, I-80126,
	Napoli, Italy.}

\author{Sergey V. Sushkov}
\email{sergey_sushkov@mail.ru}
\affiliation{Institute of Physics, Kazan Federal University, Kremlevskaya Street 16a, Kazan 420008, Russia.}

\begin{abstract}
Adopting Noether point symmetries, we classify and integrate dynamical systems coming from Horndeski cosmologies. The method is particularly  effective both to select the form of Horndeski models and to derive exact cosmological solutions. Starting from the Lagrangians selected by the Noether symmetries, it is possible to derive several modified theories of gravity like $f(R)$ gravity, Brans-Dicke gravity,  string inspired gravity and so on.  In any case, exact solutions are found out.
\end{abstract}

\maketitle
\flushbottom
\section{Introduction}

The $\Lambda$-Cold Dark Matter Model ($\Lambda$CDM) can be considered the cosmological standard model supported by  the majority of the cosmological observations. Indeed, type Ia Supernovae, galaxy clustering, Cosmic Microwave Background  Radiation, and other observational tests, all confirm a coherent  snapshot where the Hubble fluid is dominated by a cosmic fluid that accelerates the Universe and a form of matter allowing the clustering of structures. These components constitute the so called cosmic {\it dark side}, i.e. dark energy and dark matter.    Despite of its great success in representing today's cosmological view of the Universe, $\Lambda$CDM model is plagued with several shortcomings that must be framed in a self-consistent cosmological model. Besides the difficulties to find  suitable candidates for dark matter particles from direct and indirect searches, to confirm (or not) the existence of supersymmetry at TeV-scales, as well as other problems \cite{Perivolaropoulos:2008ud,Sahni:2014ooa,Bull:2015stt}, the most significant one, is the tiny value of the cosmological constant \cite{Weinberg:1988cp,Martin:2012bt}.

The inability of General Relativity (GR), together with the $\Lambda$CDM model,  to constitute a complete theory capable of describing the gravitational interactions at all scales led the scientific community to pursue  new approaches by which GR should be modified or extended at infrared and ultraviolet scales. Many of the proposed alternatives  are motivated by the necessity of fitting  dark sector issues. Several theories \cite{Nojiri:2006ri,Copeland:2006wr,Durrer:2007re, Capozziello:2011et, Nojiri:2017ncd} with extra degrees of freedom propagated by scalar fields (quintessence, k-essence, kinetic braiding), as well as geometric extensions of GR, like $f(R)$ gravity \cite{Capozziello:2002rd} or $f(T)$ teleparallel-gravity \cite{Cai:2015emx}, have been suggested, during the last two decades to address  the observed accelerating expansion of the  Universe as well us the clustering of structures \cite{annalen,vesna}. In 1974, Horndeski developed  \cite{Horndeski} the most general scalar-tensor theory (with a single scalar field) with second order field equations\footnote{Theories with higher than second order equations of motion are, in most cases, plagued by the so called Ostrogradski instability and thus give rise to ghost degrees of freedom.}. In \cite{Deffayet:2009wt,Deffayet:2011gz}, the  Horndeski theory has been reconsidered according to a  generalization of  the covariant galileon models, already proposed in \cite{Nicolis:2008in}, as the decoupling limit of the graviton in the Dvali-Gabadadze-Porrati  model.

Starting from the previous approach, a lot of progress has been done and the Horndeski theory can now be considered as a general theory from which several modified theories of gravity  can be recovered. Scalar-tensor models, such as Brans-Dicke, k-essence, kinetic braiding, as well as the scalar-tensor analogue of $f(R)$ gravity, are nothing else but special cases of the Horndeski action. Apart from cosmology, significant progress has been done at smaller scales in this theory. Specifically, charged black hole solutions have been studied in the context of this theory \cite{Babichev:2016rlq,Babichev:2015rva,Babichev:2016fbg,Babichev:2017guv, Bhattacharya:2015iha}; numerical simulations for neutron stars in specific subclasses of this theory have also been developed \cite{Maselli:2016gxk,Cisterna:2015yla}. Recently, in  \cite{Nunes:2017bwb}, the authors reviewed the Horndeski cosmologies that have asymptotically de Sitter critical point. In \cite{Kobayashi:2011nu},  generalized galileons are considered as the most general framework to develop single-field inflationary models. Moreover, in \cite{Tsujikawa:2014mba},  the author proves that Horndeski theory is part of the effective field theory of cosmological perturbations, which is also a useful framework to develop inflation. Finally, in \cite{Salzano:2016udu}, the authors considered possible breaking of the Vainshtein mechanism, in a generalized Horndeski theory (or  generalized galileon model), and they claim that such a breaking  could be  responsible for  gravitational effects attributed to dark matter.

In order to tackle the Cosmological Constant problem, or the evolution of cosmological vacuum energy,  new degrees of freedom for the gravitational field have to be considered. This can be achieved  by introducing in the theory non-minimally coupled scalar field, together with higher order derivatives, in the framework of the Horndeski theory. Even though a lot of work has been done on the fact that  scalar fields may or may not couple with matter, the predominant opinion is that matter-fields do couple, with the field being ``screened'' (=hidden) at small scales. This  screening mechanisms could solve several problems and, among them, the Cosmological Constant problem. Three such mechanisms are known; the chameleon, the symmetron and the Vainshtein mechanism  \cite{Khoury:2010xi,deRham:2012az}. Although, all of them emerge in scalar-tensor theories, the latter is explicitly seen in massive gravity, in galileon and thus in Horndeski theory. Simply,  this mechanism  ``hides'' the effects of the non-linear kinetic terms inside the so called Vainshtein radius, allowing them to play an important role only at large infrared scales, that is in  cosmology  as pointed out in \cite{Babichev:2013usa}.

The Horndeski theory contains a lot of degrees of freedom  encoded in the arbitrary functions of the action: $G_i(\phi,X)$, where $i=2,...,5$, $\phi$ is the scalar field and $X=-1/2(\partial_{\mu}\phi\partial^{\mu}\phi)$ its kinetic term. The aim of this paper is to classify the Horndeski models according to  the {\it Noether Symmetry Approach} \cite{cimento}.
This method helps to find exact solutions for a given theory, once a symmetry exists. Besides, the existence of a symmetry ``selects" the integrable form of a model in a given class of theories. Finally, the symmetries of a theory are always connected to conserved quatities, according to the Noether's Theorem, and thus observables.
Here, we  classify the Horndeski models  according to the specific forms of  functions $G_i(\phi,X)$ assuming the only  criterion that the field equations are invariant under Noether point symmetries. Specifically, we apply the Noether symmetry approach as a selection criterion to determine the form of the arbitrary functions $G_i$. See also \cite{epjc} for a detailed discussion. Recently, a similar article has appeared in the literature \cite{Dimakis:2017zdu}; however, the similarity with this one is only the fact that they discuss a general family of scalar-tensor Lagrangian. They study a part of the cosmological Horndeski Lagrangian and their results are very interesting, however, we consider the whole Horndeski action.

The paper is organized as follows: In Sec. 2, a summary of Horndeski gravity and cosmology is presented. In Sec. 3, we present the Noether Symmetry Approach specifically for the Horndeski cosmology. Sec. 4 is a discussion on how specific modified theories of gravity can be recovered in this general scheme. In particular, we discuss Brans-Dicke gravity, $f(R)$ gravity, cubic galileon gravity, string motivated gravity and models with non-minimal derivative coupling. In any of these models, the form of Lagrangian is fixed by the existence of  Noether symmetry and exact solutions are derived. Discussion and conclusions are presented in Sec. 5.

\section{The Horndeski Gravity}

As we already mentioned, Horndeski wrote  the most general scalar-tensor theory of gravity with second order derivatives in the action, but with second order equations of motion. The action is given by the sum of the integrals of four different Lagrangians, i.e.
\begin{equation}\label{eq1}
\mathcal{S}_{Horndeski} = \sum _{i=2} ^5 \int d^4 x \sqrt{-g}\mathcal{L}_i\,,
\end{equation}
where
\begin{eqnarray}\label{eq2}
\mathcal{L}_2 &=& G_2\left(\phi , X\right) \,, \\
\mathcal{L}_3 &=& - G_3\left(\phi , X\right)\square \phi \,, \\
\mathcal{L}_4 &=& G_4\left(\phi , X\right)R + G_{4X} \left[\left( \square \phi\right)^2 -\left( \nabla _{\mu}\nabla _{\nu}\phi\right)^2\right] \,, \\
\mathcal{L}_5 &=& G_5\left(\phi , X\right) G_{\mu\nu} \nabla ^{\mu}\nabla ^{\nu} \phi - \frac{1}{6}G_{5X}\left[\left(\square \phi\right)^3- 3\square \phi \left( \nabla _{\mu}\nabla _{\nu}\phi\right)^2 + 2 \left( \nabla _{\mu}\nabla _{\nu}\phi\right)^3\right]\,.
\end{eqnarray}
The functions $G_2\left(\phi, X\right),\, G_3 \left(\phi, X\right),\, G_4\left(\phi, X\right)$ and $G_5\left(\phi, X\right)$ are arbitrary functions of the scalar field $\phi$ and its kinetic term $X=- \frac{1}{2}\left(\nabla \phi\right)^2 =-\frac{1}{2} \nabla^{\mu}\phi \nabla_{\mu}\phi$. In addition, $G_{iX}$ is the derivative of $G_i$ with respect to $X$, $R$ is the Ricci scalar,  ${\displaystyle G_{\mu\nu}=R_{\mu\nu}-\frac{1}{2}g_{\mu\nu}R}$ is the Einstein tensor, and the remaining kinetic terms are
\begin{eqnarray}\label{eq3}
\square \phi &=& g^{\mu\nu}\nabla_{\mu}\nabla_{\nu}\phi\,,\\
\left(\nabla _{\mu} \nabla _{\nu} \phi \right) ^2 &=& \nabla ^{\mu} \nabla ^{\nu} \phi \nabla _{\mu} \nabla _{\nu} \phi \,,\\
\left(\nabla _{\mu} \nabla _{\nu} \phi \right) ^3 &=& \nabla _{\mu} \nabla _{\nu} \phi \nabla ^{\nu} \nabla ^{\lambda} \phi \nabla _{\lambda} \nabla ^{\mu} \phi \,.
\end{eqnarray}
If we vary the action with respect to the metric and the scalar field,  we get the field equations for the Horndeski theory \cite{Kobayashi:2011nu}. The variation is 
\begin{equation}\label{eq4}
\delta \mathcal{S} = \delta \left( \sqrt{-g} \sum _{i=2}^{5} \mathcal{L}_i \right)= \sqrt{-g} \left[ \sum _{i=2}^{5} \mathcal{G}_{\mu\nu}^i \delta g^{\mu\nu} + \sum _{i=2}^{5}\left(P_{\phi}^i- \nabla^{\mu}J_{\mu}^i\right)\delta \phi\right] + \text{total derivatives},
\end{equation}
and thus the equations of motion are given by
\begin{equation}
\sum_{i=2}^{5} \mathcal{G}_{\mu\nu}^i = 0 \,,\quad \nabla^{\mu}\left( \sum_{i=2}^5 J_{\mu}^i\right) = \sum _{i=2}^{5} P_{\phi}^i\,,
\end{equation}
for the metric and the scalar field respectively. The components are 
\begin{subequations}
\begin{align}
P_{\phi}^{2} &= G_{2\phi}\,,\\
P_{\phi}^{3} &= \nabla_{\mu}G_{3\phi}\nabla^{\mu} \phi \,,\\
P_{\phi}^{4} &= G_{4\phi}R + G_{4\phi X} \left[ (\square \phi)^2 - (\nabla_{\mu}\nabla_{\nu}\phi)^2\right]\,,\\
P_{\phi}^{5} &= -\nabla_{\mu}G_{5\phi} G^{\mu\nu}\nabla_{\nu}\phi - \frac{1}{6}G_{5\phi X}\left[(\square \phi)^3 - 3 \square \phi (\nabla_{\mu}\nabla_{\nu} \phi)^2 + 2 (\nabla_{\mu}\nabla_{\nu}\phi)^3 \right]\,,
\end{align}
\end{subequations}
and
\begin{subequations}
\begin{align}
J_{\mu}^{2} &= -\mathcal{L}_{2X}\nabla_{\mu}\phi \,,\\
J_{\mu}^{3} &= -\mathcal{L}_{3X}\nabla_{\mu}\phi + G_{3X} \nabla_{\mu}X + 2 G_{3\phi} \nabla_{\mu} \phi \,,\\
J_{\mu}^{4} &= - \mathcal{L}_{4X}\nabla_{\mu} \phi +2 G_{4X}R_{\mu\nu}\nabla^{\nu} \phi  - 2 G_{4XX}\left(\square \phi \nabla_{\mu}X - \nabla^{\nu} X \nabla_{\mu}\nabla_{\nu} \phi \right) \nonumber \\
&-2 G_{4\phi X} (\square \phi \nabla_{\mu}\phi + \nabla_{\mu}X) \,,\\
J_{\mu}^{5} &= -\mathcal{L}_{5X}\nabla_{\mu}\phi - 2 G_{5\phi}G_{\mu\nu} \nabla^{\nu} \phi - \nonumber \\
&-G_{5X}\left[ G_{\mu\nu}\nabla^{\nu} X + R_{\mu\nu}\square \phi \nabla^{\nu} \phi - R_{\nu \lambda} \nabla^{\nu} \phi \nabla^{\lambda}\nabla_{\mu} \phi - R_{\alpha \mu \beta \nu}\nabla^{\nu}\phi \nabla^{\alpha} \nabla^{\beta}\phi\right] + \nonumber \\
&+G_{5XX} \lbrace \frac{1}{2}\nabla_{\mu}X \left[(\square \phi)^2 - (\nabla_{\alpha}\nabla_{\beta}\phi)^2 \right]- \nabla_{\nu}X\left(\square \phi \nabla_{\mu}\nabla^{\nu}\phi - \nabla_{\alpha}\nabla_{\mu}\phi\nabla^{\alpha}\nabla^{\nu}\phi\right)\rbrace + \nonumber \\
&+G_{5\phi X} \lbrace \frac{1}{2}\nabla_{\mu}\phi \left[(\square \phi)^2 - (\nabla_{\alpha}\nabla_{\beta}\phi)^2 \right] + \square \phi \nabla_{\mu}X -\nabla^{\nu}X \nabla_{\nu}\nabla_{\mu}\phi  \rbrace\,,
\end{align}
\end{subequations}
as well as
\begin{subequations}
\begin{align}
\mathcal{G}_{\mu\nu}^2 &= - \frac{1}{2} G_{2X} \nabla_\mu \phi \nabla _{\nu} \phi - \frac{1}{2}G_2 g_{\mu\nu}\,,\\
\mathcal{G}_{\mu\nu}^3 &= \frac{1}{2}G_{3X} \square \phi \nabla_{\mu} \phi \nabla_{\nu}\phi + \nabla _{(\mu} G_3 \nabla _{\nu)} \phi - \frac{1}{2}g_{\mu\nu} \nabla_{\lambda} G_3 \nabla ^{\lambda} \phi\,,\\
\mathcal{G}_{\mu\nu}^4 &= G_4 G_{\mu\nu} - \frac{1}{2} G_{4X} R \nabla_{\mu}\phi \nabla_{\nu} \phi - \frac{1}{2} G_{4XX} \left[ (\square \phi)^2 - (\nabla_{\alpha}\nabla_{\beta} \phi)^2\right]\nabla_{\mu}\phi \nabla_{\nu}\phi - \nonumber \\
& - G_{4X} \square \phi \nabla _{\mu}\nabla _{\nu} \phi + G_{4X} \nabla_{\lambda}\nabla_{\mu}\phi \nabla ^{\lambda}\nabla_{\nu} \phi  + 2 \nabla_{\lambda} G_{4X} \nabla^{\lambda} \nabla_{(\mu}\phi \nabla_{\nu)} \phi  - \nonumber \\
&- \nabla_{\lambda}G_{4X}\nabla^{\lambda}\phi \nabla_{\mu}\nabla_{\nu}\phi + g_{\mu\nu} \left(G_{4\phi} \square \phi - 2 X G_{4\phi\phi} \right) + \nonumber \\
&+ g_{\mu\nu} \lbrace -2 G_{4 \phi X} \nabla_{\alpha} \nabla_{\beta} \phi \nabla^{\alpha} \phi\nabla^{\beta}\phi + G_{4XX}\nabla_{\alpha}\nabla_{\lambda}\phi \nabla_{\beta}\nabla^{\lambda}\phi \nabla ^{\alpha}\phi \nabla^{\beta}\phi + \nonumber \\
&+ \frac{1}{2} G_{4X} \left[(\square \phi)^2-(\nabla_{\alpha}\nabla_{\beta} \phi)^2 \right] \rbrace +   2 \Big[ G_{4X} R_{\lambda(\mu} \nabla_{\nu)}\phi \nabla^{\lambda}\phi - \nabla_{(\mu} G_{4X} \nabla_{\nu)} \phi \square \phi \Big] -\nonumber \\
&-  g_{\mu\nu} \left[ G_{4X}R^{\alpha\beta}\nabla_{\alpha}\phi\nabla_{\beta}\phi - \nabla_{\lambda}G_{4X}\nabla^{\lambda}\phi \square \phi\right]+  G_{4X}R_{\mu\alpha\nu\beta}\nabla^{\alpha}\phi\nabla^{\beta}\phi - \nonumber \\
&- G_{4\phi}\nabla_{\mu}\nabla_{\nu}\phi - G_{4\phi\phi}\nabla_{\mu}\phi\nabla_{\nu}\phi + 2 G_{4\phi X} \nabla^{\lambda} \phi \nabla_{\lambda}\nabla_{(\mu}\phi\nabla_{\nu)}\phi -\nonumber \\
&- G_{4XX}\nabla^{\alpha}\phi \nabla_{\alpha}\nabla_{\mu}\phi\nabla^{\beta} \phi \nabla_{\beta}\nabla_{\nu}\phi \, ,\\
\mathcal{G}_{\mu\nu}^5 &=G_{5X} R_{\alpha \beta} \nabla^{\alpha} \phi \nabla^{\beta} \nabla_{(\mu}\phi \nabla_{\nu)} \phi - G_{5X} R_{\alpha ( \mu} \nabla_{\nu)} \phi \nabla^{\alpha} \phi \square \phi -  \frac{1}{2} G_{5X}R_{\mu\alpha\nu\beta}\nabla^{\alpha}\phi \nabla^{\beta}\phi \square\phi -\nonumber  \\
&- \frac{1}{2} G_{5X}R_{\alpha \beta}\nabla^{\alpha}\phi \nabla^{\beta}\phi \nabla_{\mu}\nabla_{\nu}\phi  +  G_{5X}R_{\alpha\lambda\beta(\mu}\nabla_{\nu)}\phi\nabla^{\lambda}\phi \nabla^{\alpha}\nabla^{\beta}\phi - \nonumber \\
&- \frac{1}{2} \nabla_{(\mu} \left[G_{5X}\nabla^{\alpha}\phi \right] \nabla_{\alpha}\nabla_{\nu)}\phi \square \phi  + \frac{1}{2} \nabla_{(\mu} \left[G_{5\phi}\nabla_{\nu)} \right] \square \phi -    \nabla_{\lambda} \left[G_{5\phi}\nabla_{(\mu} \phi\right] \nabla_{\nu)} \nabla^{\lambda}\phi + \nonumber \\
&+ \frac{1}{2} \left[ \nabla_{\lambda }\left(G_{5\phi} \nabla^{\lambda}\phi \right)- \nabla_{\alpha}\left(G_{5X}\nabla_{\beta} \phi \right) \nabla^{\alpha}\nabla^{\beta}\phi\right] \nabla_{\mu}\nabla_{\nu}\phi +  \nabla^{\alpha} G_5 \nabla^{\beta} \phi R_{\alpha (\mu\nu)\beta} + \nonumber \\
&+ \frac{1}{2}\nabla_{(\mu}G_{5X}\nabla_{\nu)} \phi -\left[ (\square \phi)^2 - (\nabla_{\alpha}\nabla_{\beta} \phi)^2\right] + G_{5X} R_{\alpha \lambda \beta (\mu}\nabla_{\nu)}\nabla^{\lambda} \phi \nabla^{\alpha}\phi \nabla^{\beta} \phi  - \nonumber \\
&- \nabla^{\lambda} G_5 R_{\lambda(\mu} \nabla_{\nu)}\phi + \nabla_{\alpha}\left[G_{5X}\nabla_{\beta}\phi \right] \nabla^{\alpha} \nabla_{(\mu}\phi \nabla^{\beta} \nabla_{\nu)}\phi  - \nabla_{(\mu}G_5G_{\nu ) \lambda}\nabla^{\lambda}\phi  -  \nonumber \\
&-  \nabla_{\beta} G_{5X}\left[ \square \phi \nabla^{\beta} \nabla_{(\mu}\phi - \nabla^{\alpha}\nabla^{\beta} \phi \nabla_{\alpha}\nabla_{(\mu}\phi \right] \nabla_{\nu)}\phi +  \nonumber \\
&+ \frac{1}{2} \nabla^{\alpha} \phi \nabla_{\alpha} G_{5X} \left[ \square \phi \nabla_{\mu} \nabla_{\nu} \phi - \nabla_{\beta}\nabla_{\mu}\phi \nabla^{\beta} \nabla_{\nu}\phi\right] + \frac{1}{2} \nabla_{\lambda}G_5G_{\mu\nu} \nabla^{\lambda}\phi - \nonumber \\
&- \frac{1}{2} G_{5X} G_{\alpha \beta} \nabla^{\alpha} \nabla^{\beta} \phi \nabla_{\mu} \phi \nabla_{\nu} \phi - \frac{1}{2} G_{5X}\square \phi \nabla_{\alpha}\nabla_{\mu}\phi \nabla^{\alpha}\nabla_{\nu}\phi + \nonumber \\
&+ \frac{1}{2}G_{5X}(\square \phi)^2 \nabla_{\mu} \nabla_{\nu} \phi +\frac{1}{12} G_{5XX} \big[ (\square \phi)^3 - 3 \square \phi (\nabla_{\alpha}\nabla_{\beta}\phi)^2 +2 (\nabla_{\alpha} \nabla_{\beta} \phi)^3 \big] \nabla_{\mu}\nabla_{\nu}\phi + \nonumber \\
&+ g_{\mu\nu} \lbrace -\frac{1}{6} G_{5X} \left[(\square \phi)^3 - 3\square \phi (\nabla_{\alpha}\nabla_{\beta}\phi)^2 + 2 (\nabla_{\alpha}\nabla_{\beta} \phi)^3 \right] + \nabla _{\alpha}G_5 R^{\alpha \beta}\nabla_{\beta} \phi - \nonumber \\
&- \frac{1}{2} \nabla_{\alpha} \left(G_{5\phi}\nabla^{\alpha} \phi \right) \square \phi + \frac{1}{2} \nabla _{\alpha} \left( G_{5\phi} \nabla_{\beta} \phi \right) \nabla^{\alpha}\nabla^{\beta} \phi - \frac{1}{2} \nabla_{\alpha} G_{5X} \nabla^{\alpha}X \square \phi +\nonumber \\
&+ \frac{1}{2} \nabla_{\alpha} G_{5X} \nabla_{\beta} X \nabla^{\alpha} \nabla^{\beta} \phi -\frac{1}{4} \nabla^{\lambda} G_{5X} \nabla_{\lambda} \phi \left[ (\square \phi)^2-(\nabla_{\alpha}\nabla_{\beta}\phi)^2\right] + \nonumber \\
&+ \frac{1}{2} G_{5X} R_{\alpha\beta} \nabla^{\alpha}\phi \nabla^{beta}\phi  \square \phi - \frac{1}{2}G_{5X}R_{\alpha \lambda\beta \rho} \nabla^{\alpha}\nabla^{\beta} \phi \nabla^{\lambda} \phi \nabla^{\rho} \phi \rbrace\,,
\end{align}
\end{subequations}
It is  easy to see that, from \eqref{eq1}, one can derive several already known models. For example, if $G_2= \frac{\omega}{\phi}X\,,\,\,G_3 = 0\,,\,\,G_4=\phi\,,$ and $G_5=0$, we obtain the Brans-Dicke theory and so on. What we will show in the rest of the paper is, how to choose the form of these functions by a geometric criterion based on the existence of  Noether point symmetries.

\subsection{The Horndeski Cosmology}
We want to study the cosmology related to the above theory, so we suppose that the spacetime is described by a  spatially flat Friedmann-Robertson-Walker (FRW) metric, which reads
\begin{equation}\label{eq5}
ds^2 = - dt^2 + a^2(t) \delta _{ij}dx^i dx^j\,.
\end{equation}
The Ricci scalar takes the form
\begin{equation}\label{eq6}
R= 6\left(\frac{\ddot{a}}{a}+\frac{\dot{a}^2}{a^2}\right)\,.
\end{equation}
It is $\phi = \phi (t)$  and thus the scalars in the Lagrangians become\footnote{This is explained if we assume that the matter fields and  scalar field inherit the isometries of the FRW spacetime.}  \begin{equation}\label{eq7}
X = \frac{1}{2} \dot{\phi}^2 \,,\,\, \square \phi = - \left(\ddot{\phi}+3 \frac{\dot{a}}{a}\dot{\phi} \right) \,,\,\, \left(\nabla _{\mu} \nabla _{\nu} \phi \right) ^2 = \ddot{\phi}^2 + 3 \frac{\dot{a}^2}{a^2}\dot{\phi}^2 \,,\,\,  \left(\nabla _{\mu} \nabla _{\nu} \phi \right) ^3 = - \ddot{\phi}^3 - \frac{3 \dot{a}^3 }{a^3} \dot{\phi}^3 \,.
\end{equation}
If we substitute all these quantities into \eqref{eq1}, the Lagrangian assumes a point-like  form
\begin{eqnarray}\label{eq8}
\mathcal{L} =  a^3 G_2 + 3 a^2 G_3 \dot{a} \dot{\phi} + a^3  G_3 \ddot{\phi}  +  6 a  G_{4X} \dot{a}^2 \dot{\phi} ^2 + 6 a G_4 \dot{a}^2 + 6 a^2 G_{4X} \dot{a} \dot{\phi}  \ddot{\phi}  + 6 a^2 G_4 \ddot{a}  + \nonumber \\
+ 3 a  G_5 \dot{a}^2 \ddot{\phi} + 6 a  G_5 \dot{a} \dot{\phi} \ddot{a}  + 3 G_5 \dot{a}^3 \dot{\phi} + G_{5X} \dot{a}^3 \dot{\phi}^3 + 3 a G_{5X} \dot{a}^2 \dot{\phi}^2 \ddot{\phi} \,.
\end{eqnarray}
As we see, there are second order derivatives in the Lagrangian. We can integrate all of them out with integration by parts, except from the term $a^3 G_3 \ddot{\phi}$. Specifically,
\begin{eqnarray}
a^3 G_3 \ddot{\phi} &=& (a^3 G_3 \dot{\phi})_{,t} - 3 a^2 G_3 \dot{a}\dot{\phi} -a^3 G_{3\phi}\dot{\phi}^2 - a^3 G_{3X} \dot{\phi}^2 \ddot{\phi}\,\nonumber \\
&=& (a^3 G_3 \dot{\phi})_{,t} - 3 a^2 G_3 \dot{a}\dot{\phi} -a^3 G_{3\phi}\dot{\phi}^2 + a^2 G_{3X} \dot{a} \dot{\phi}^3 +\frac{1}{3} a^3 G_{3X\phi} \dot{\phi}^4 + \frac{1}{3} a^3 G_{3XX} \dot{\phi}^4 \ddot{\phi}\,, \nonumber
\end{eqnarray}
and it goes on like this, since $G_3$ depends on $X(t)$ and $\dot{X}(t) = \dot{\phi} \ddot{\phi}$. Hence, if we want the Lagrangian to be canonical and to depend  only on first derivatives of the variables of the configuration space\footnote{In our case the configuration space is the minisuperspace   $\mathcal{Q} = \{a,\phi \}$ and the  tangent space is $\mathcal{TQ} = \{a,\dot{a},\phi,\dot{\phi}\}$.}, we have to choose where to stop and just set one derivative of $G_3$ over $X$ equal to zero. We  choose to set
\begin{equation}\label{G3XX}
G_{3XX} = 0 \Rightarrow G_3(\phi,X) = g(\phi) X + h(\phi) \,.
\end{equation}
This choice  seems arbitrary, but also with this limitation, it is possible to  realize the most of  scalar-tensor theories studied in literature, such as kinetic braiding, cubic galileons and others  containing interaction terms like  $\sim \nabla_{\mu}\phi \nabla^{\mu} \phi \square \phi$.  Finally, the Lagrangian \eqref{eq8} becomes
\begin{eqnarray}\label{eq9}
\mathcal{L} = a^3 G_2  + a^2 g(\phi ) \dot{a} \dot{\phi}^3 - \frac{1}{6} a^3  g'(\phi ) \dot{\phi}^4 - a^3  h'(\phi ) \dot{\phi}^2 - 6 a G_4 \dot{a}^2 - 6 a^2  G_{4\phi} \dot{a} \dot{\phi} + \nonumber \\
+ 3 a \left(2 G_{4X}-G_{5\phi}\right)\dot{a}^2 \dot{\phi}^2  + G_{5X} \dot{a}^3 \dot{\phi}^3\,.
\end{eqnarray}
The Euler-Lagrange equations
\begin{equation}\label{eq10}
\frac{d}{dt}\left(\frac{\partial \mathcal{L}}{\partial \dot{a}}\right)-\frac{\partial \mathcal{L}}{\partial a} = 0\,,
\qquad
\frac{d}{dt}\left(\frac{\partial \mathcal{L}}{\partial \dot{\phi}}\right)-\frac{\partial \mathcal{L}}{\partial \phi} = 0\,,
\end{equation} 
and the energy condition
\begin{equation}
E_\mathcal{L}=\frac{\partial\mathcal{L}}{\partial\dot{a}}\dot{a}+\frac{\partial\mathcal{L}}{\partial\dot{\phi}}\dot{\phi}-\mathcal{L}=0\,,       \label{4.37}
\end{equation}
 constitute  the dynamical system derived from the Lagrangian \eqref{eq9}. We do not find necessary to include them in their general form since they can be easily derived from the Lagrangian \eqref{eq9}. We will derive them for the specific cases that we are going to discuss below.

\section{ The Noether  Symmetry Approach }

\subsection{The point symmetries}
Let us see how a differential equation behaves under the action of a point transformation. A given  differential equation has the form $D=D(t, q^i) = 0$, where $t$ is the independent variable and $\{q^i : i=1,2,...,n\}$ are the configurations. Suppose that a one parameter point transformation is expressed by
\begin{equation}\label{eq12}
\bar{t} = Z (t, q^i, \delta) \,,\,\, \bar{q}^i = \Gamma ^i (t, q^i, \delta)\,,
\end{equation}
and hence the generator of transformations is given by
\begin{equation}\label{eq13}
\X = \xi(t, q^i) \partial _t + \eta ^i (t, q^i)\partial _i\,,
\end{equation}
where
\begin{equation}\label{eq14}
\xi (t, q^i) = \frac{\partial Z (t, q^i, \delta)}{\partial \delta}\mid _{\delta \rightarrow 0}\,,\,\, \eta^i(t,q^i)= \frac{\partial \Gamma ^i (t,q^i,\delta)}{\partial \delta}\mid _{\delta \rightarrow 0}\,.
\end{equation}
In this case, the $n^{th}$ prolongation of the generator \eqref{eq13} is
\begin{equation}\label{eq15}
\X^{[n]} = \X + \eta ^i _{[1]} \partial _{\dot{q}^i} + ... + \eta ^i _{[n]} \partial _{q^{(n) i}}\,,
\end{equation}
where
$\eta ^i _{t} = D_t \eta^i - q^i D_t \xi \,,$
and $D_t = \frac{\partial}{\partial t} + \dot{q}^i \frac{\partial }{\partial q^i}\,.$

Let \eqref{eq13} be the generator of an infinitesimal transformation and $L=L(t,q^i,\dot{q}^i)$ be a Lagrangian of a dynamical system. Then the Euler-Lagrange equations
\begin{equation}\label{eq17}
E_i (L) =0 \quad  \Rightarrow \quad \frac{d}{dt} \frac{\partial L}{\partial \dot{q}^i}- \frac{\partial L}{\partial q^i} = 0\,,
\end{equation}
are invariant under the transformation iff there exists a function $f=f(t,q^i)$ such that the following condition holds
\begin{equation}\label{eq18}
\X^{[1]} L + L \frac{d\xi}{dt}= \frac{df}{dt}\,,
\end{equation}
where $\X^{[1]}$ is the first prolongation of the generating vector \eqref{eq13}. This method was first introduced in \cite{Basilakos:2011rx}. In this case,  the generator is a Noether symmetry of the dynamical system described by $L$. For any Noether symmetry there exists a function
\begin{equation}\label{eq20}
I = \xi \left(\dot{q}^i\frac{\partial L}{\partial \dot{q}^i}-L\right) - \eta ^i \frac{\partial L}{\partial q^i} + f\,,
\end{equation}
which is a first integral, i.e. $\frac{dI}{dt} = 0$ of the equations of motion \eqref{eq17}.
In the   case of above Horndeski Lagrangian \eqref{eq9}, it is
\begin{equation}
\X^{[1]} = \X + \eta_{a} ^{[1]}\partial _{\dot{a}}+ \eta_{\phi} ^{[1]} \partial _{\dot{\phi}}\,,
\end{equation}
and
\begin{eqnarray}
\eta_a ^{[1]} &=& \left( \partial _t \eta_{a} + \dot{a} \partial _a \eta_{a} + \dot{\phi} \partial_{\phi} \eta_{a} - \dot{a} \partial _t \xi - \dot{a}^2 \partial _a \xi - \dot{a} \dot{\phi}\partial _{\phi}\xi \right)\,,\\
\eta_{\phi} ^{[1]} &=& \left( \partial _t \eta_{\phi} + \dot{a} \partial _a \eta_{\phi} + \dot{\phi} \partial_{\phi} \eta_{\phi} - \dot{\phi} \partial _t \xi - \dot{\phi}^2 \partial _{\phi} \xi - \dot{a} \dot{\phi}\partial _{a}\xi \right)\,,
\end{eqnarray}
 then the Noether integral is
 \begin{equation}
I = f- \eta_{a} \frac{\partial L}{\partial \dot{a}}-\eta_{\phi} \frac{\partial L}{\partial \dot{\phi}}\,. 
\end{equation}
In what follows, we will give explicit examples for the above considerations. Further details on the method can be found  in \cite{sebastian}.

\subsection{Noether Symmetries in Horndeski Cosmology}
As we said, the configuration space of the Lagrangian \eqref{eq9} is ${\cal Q}=\{a,\phi\}$ and the independent variable is the cosmic time $t$. The generator of an infinitesimal transformation  is
\begin{equation}\label{eq19}
\X = \xi \left(t,a,\phi \right) \partial _t + \eta_{a} \left(t,a,\phi \right) \partial _a + \eta_{\phi} \left(t,a,\phi \right) \partial _{\phi}\,.
\end{equation}
By applying \eqref{eq18} to \eqref{eq9}, we get a system of 28 equations for the coefficients of the Noether vector $\xi (t,a,\phi) \,,$  $\eta_{a} (t,a,\phi)\,,$ $\eta_{\phi}(t,a,\phi)\,,$ $f(t,a,\phi)$ and the arbitrary functions of the Lagrangian $G_2(\phi,X), G_3(\phi,X), G_4(\phi,X), G_5(\phi,X)$, which, of course, are not all each other  independent (see also \cite{cimento,sebastian}). A comment  is necessary at this point.  If we consider  given  forms for  the unknown functions of the Lagrangian, i.e.  the $G_i$, as it has been done in  other papers, see for example
 \cite{Dimakis:2017zdu,Paliathanasis:2014rja,Dimakis:2017kwx,Giacomini:2017yuk}, we can specify in  detail all the functions, as well as the Noether vector coefficients. What we are doing here is to consider the most general Horndeski Lagrangian and try to constrain its unknown functions and, at  the same time, to find out  symmetries in the most general way. Clearly, particular models are recovered by specific choices of the above functions,  as we will show below with some examples. 
 
It is straightforward to notice that the Noether vector takes immediately the form
\begin{equation}\label{eq22}
\X = (\xi _1 t + \xi_2)\partial_t + \eta_{a}(a) \partial_a +(\xi_1 \phi + \phi_1) \partial_{\phi} \,,
\end{equation}
with $\xi_1, \,\xi_2,\,\phi_1$ being integration constants. In addition, the function $f$ of Eq. \eqref{eq18} is forced to be a constant, $f(t,a,\phi) = f_1$.

Now, depending on whether the function $g(\phi)$ in Eq. \eqref{G3XX} vanishes or not, there are different solutions. In the class of solutions with $g(\phi) \neq 0$,  the Noether vector, and specifically the $\eta_a$ coefficient, becomes $\eta_a (a)  = \alpha_1 a$. In the other case, where $g(\phi) =0 $, we get $\eta_a(a) = \frac{1}{3}(\alpha_1 + 2 \xi_1)a$. It might seem that a redefinition of the constants would equate the two cases, but this is not the case. As we show in Table \ref{tab:sum}, the Horndeski functions take different forms.

The following graph summarizes the 10 different symmetry classes we get depending on the values of the constants. By changing $\xi_1$ and $\alpha_1$, the form of  symmetry, i.e. the Noether vector, changes in a straightforward way. In the graph, any different case is assigned to a capital letter; the Horndeski functions, for each case, are given in Table \ref{tab:sum}. The cases {\bf A}, {\bf J} and {\bf B}, {\bf I} coincide by redefining the constants and by setting $c_2 = 0$ in {\bf A} and {\bf B}. However, the other  cases are different.

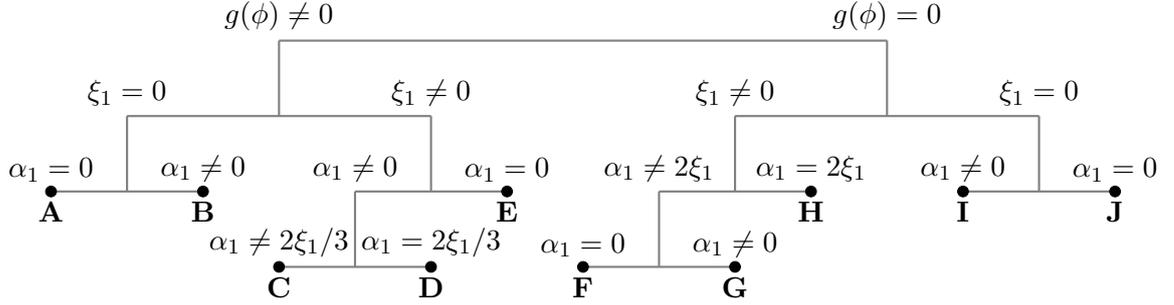
\begin{center}

\begin{tikzpicture}
\draw[gray, thick] (-4,0) -- (4,0);
\draw[gray, thick] (-4,0) -- (-4,-1);
\draw[gray, thick] (-6,-1) -- (-2,-1);
\draw[gray, thick] (-6,-1) -- (-6,-2);
\draw[gray, thick] (-7,-2) -- (-5,-2);
\draw[gray, thick] (4,0) -- (4,-1);
\draw[gray, thick] (2,-1) -- (6,-1);
\draw[gray, thick] (2,-1) -- (2,-2);
\draw[gray, thick] (1,-2) -- (3,-2);
\draw[gray, thick] (1,-2) -- (1,-3);
\draw[gray, thick] (0,-3) -- (2,-3);
\draw[gray, thick] (-2,-1) -- (-2,-2);
\draw[gray, thick] (-3,-2) -- (-1,-2);
\draw[gray, thick] (-3,-2) -- (-3,-3);
\draw[gray, thick] (-4,-3) -- (-2,-3);
\draw[gray, thick] (6,-1) -- (6,-2);
\draw[gray, thick] (5,-2) -- (7,-2);
\filldraw[black] (-4,0)  node[anchor=south] {$g(\phi) \neq 0$};
\filldraw[black] (4,0)  node[anchor=south] {$g(\phi) = 0$};
\filldraw[black] (-6,-1)  node[anchor=south] {$\xi_1 = 0$};
\filldraw[black] (-2,-1)  node[anchor=south] {$\xi_1 \neq 0$};
\filldraw[black] (2,-1)  node[anchor=south] {$\xi_1 \neq 0$};
\filldraw[black] (6,-1)  node[anchor=south] {$\xi_1 = 0$};
\filldraw[black] (5,-2) circle (2pt) node[anchor=south] {$\alpha_1 \neq 0$};
\filldraw[black] (7,-2) circle (2pt) node[anchor=south] {$\alpha_1 = 0$};
\filldraw[black] (-7,-2) circle (2pt) node[anchor=south] {$\alpha_1 = 0$};
\filldraw[black] (-5,-2) circle (2pt) node[anchor=south] {$\alpha_1 \neq 0$};
\filldraw[black] (-3,-2)  node[anchor=south] {$\alpha_1 \neq 0$};
\filldraw[black] (-4,-3) circle (2pt) node[anchor=south] {$\alpha_1 \neq 2 \xi_1/3$};
\filldraw[black] (-2,-3) circle (2pt) node[anchor=south] {$\alpha_1 = 2 \xi_1/3$};
\filldraw[black] (-1,-2) circle (2pt) node[anchor=south] {$\alpha_1 = 0$};
\filldraw[black] (1,-2) node[anchor=south] {$\alpha_1 \neq 2 \xi_1$};
\filldraw[black] (3,-2) circle (2pt) node[anchor=south] {$\alpha_1 = 2 \xi_1$};
\filldraw[black] (0,-3) circle (2pt) node[anchor=south] {$\alpha_1 = 0$};
\filldraw[black] (2,-3) circle (2pt) node[anchor=south] {$\alpha_1 \neq 0 $};
\filldraw[black] (-7,-2)  node[anchor=north] {\textbf{A}};
\filldraw[black] (-5,-2)  node[anchor=north] {\textbf{B}};
\filldraw[black] (-4,-3)  node[anchor=north] {\textbf{C}};
\filldraw[black] (-2,-3)  node[anchor=north] {\textbf{D}};
\filldraw[black] (-1,-2)  node[anchor=north] {\textbf{E}};
\filldraw[black] (0,-3)  node[anchor=north] {\textbf{F}};
\filldraw[black] (2,-3)  node[anchor=north] {\textbf{G}};
\filldraw[black] (3,-2)  node[anchor=north] {\textbf{H}};
\filldraw[black] (5,-2)  node[anchor=north] {\textbf{I}};
\filldraw[black] (7,-2)  node[anchor=north] {\textbf{J}};
\end{tikzpicture}
\end{center}

\begin{table}[!ht]
  \centering
  \resizebox{15cm}{!}{
  \begin{tabular}{r|c|c|c|clcl}
    \hline
    \hline
 & $\mathbf{G_2(\phi,X) }$ & $\mathbf{G_3(\phi,X) } $ & $\mathbf{G_4(\phi,X) } $ & $\mathbf{G_5(\phi,X) } $ \\ 
 \hline 
\textbf{A} & $g_2(X)$ & $c_1 + c_2 X+ c_3 \phi$  & $g_4(X)$ & $g_5(X)+c_4 \phi$  \\ 
\hline
 \textbf{B} & $ e^{-\frac{3 \alpha _1 \phi }{\phi _1}} g_2(X)$ & $ c_1 + e^{-\frac{3 \alpha _1 \phi }{\phi _1}} \left(c_2 X-\frac{c_3 \phi _1}{3 \alpha _1}\right)$ & $e^{-\frac{3 \alpha _1 \phi }{\phi _1}} g_4(X)$ & $c_4-\frac{\phi _1 e^{-\frac{3 \alpha _1 \phi }{\phi _1}} g_5(X)}{3 \alpha _1}$ \\ 
\hline 
 \textbf{C} & $g_2(X) \left(\xi_1 \phi +\phi_1\right)^{-\frac{3 \alpha _1}{\xi_1} - 1}$ & $\left(\xi _1 \phi +\phi _1\right){}^{-\frac{3 \alpha _1}{\xi _1}} \left(c_1 X-\frac{c_2}{3 \alpha _1}\right)+c_3$ & $g_4(X) \left(\xi _1 \phi +\phi _1\right){}^{1-\frac{3 \alpha _1}{\xi _1}}$ & $c_4+\frac{g_5(X) \left(\xi _1 \phi +\phi _1\right){}^{2-\frac{3 \alpha _1}{\xi _1}}}{2 \xi _1-3 \alpha _1}$\\ 
 \hline
 \textbf{D} & $\frac{g_2(X)}{\left(\xi _1 \phi +\phi _1\right){}^3}$ & $c_1-\frac{c_2-2 c_3 \xi _1 X}{2 \xi _1 \left(\xi _1 \phi +\phi _1\right){}^2}$ & $\frac{g_4(X)}{\xi _1 \phi +\phi _1}$ & $\frac{c_4 \ln \left(\xi _1 \phi +\phi _1\right)}{\xi _1}+g_5(X)+c_4 $ \\ 
 \hline
 \textbf{E} & $\frac{g_2(X)}{\xi _1 \phi +\phi _1}$ & $\frac{c_1 \ln \left(\xi _1 \phi +\phi _1\right)}{\xi _1}+c_2 X +c_3$ & $\left(\xi _1 \phi +\phi _1\right)g_4(X)$ & $c_4+\frac{g_5(X) \left(\xi _1 \phi +\phi _1\right){}^2}{2 \xi _1}$ \\  
 \hline
 \textbf{F} & $\frac{g_2(X)}{\left(\xi _1 \phi +\phi _1\right){}^3}$ & $c_1-\frac{c_2}{2 \xi _1 \left(\xi _1 \phi +\phi _1\right){}^2}$ & $\frac{g_4(X)}{\left(\xi _1 \phi +\phi _1\right)}$ & $\frac{c_3 \ln \left(\xi _1 \phi +\phi _1\right)}{\xi _1}+g_5(X)$ \\  
 \hline
 \textbf{G} &$ g_2(X) \left(\xi _1 \phi +\phi _1\right){}^{-\frac{\alpha _1}{\xi _1}-3}$ &$c_1-\frac{c_2 \left(\xi _1 \phi +\phi _1\right){}^{-\frac{\alpha _1}{\xi _1}-2}}{\alpha _1+2 \xi _1}$ & $g_4(X) \left(\xi _1 \phi +\phi _1\right){}^{-\frac{\alpha _1}{\xi _1}-1}$ & $c_3-\frac{g_5(X) \left(\xi _1 \phi +\phi _1\right){}^{-\frac{\alpha _1}{\xi _1}}}{\alpha _1} $\\  
 \hline
 \textbf{H} & $\frac{g_2(X)}{\xi_1 \phi +\phi_1}$ & $c_1 + \frac{c_2 \log \left(\xi _1 \phi +\phi _1\right)}{\xi _1}$ & $\frac{g_4(X) \left(\xi _1 \phi +\phi _1\right)}{\xi _1}$ & $c_3+\frac{g_5(X) \left(\xi _1 \phi +\phi _1\right){}^2}{2 \xi _1}$ \\  
 \hline
 \textbf{I} & $e^{-\frac{\alpha _1 \phi }{\phi _1}} g_2(X)$ & $c_1-\frac{c_2 \phi _1 e^{-\frac{\alpha _1 \phi }{\phi _1}}}{\alpha _1}$ & $e^{-\frac{\alpha _1 \phi }{\phi _1}} g_4(X)$ & $c_3-\frac{\phi _1 e^{-\frac{\alpha _1 \phi }{\phi _1}} g_5(X)}{\alpha _1}$ \\  
 \hline
 \textbf{J} & $g_2(X)$ & $c_1 + c_2 \phi$ & $g_4(X)$ & $c_3 \phi +g_5(X)$\\ 
   \hline
    \hline
  \end{tabular}
  }
  \caption{We summarize the Horndeski functions with respect to the Noether Symmetries; $g_i(X)$ are arbitrary functions of $X$, the kinetic term; $c_i$ are arbitrary constants and $\xi_1,\,\phi_1$ and $\alpha_1$ are the constants coming from the Noether vector. }
  \label{tab:sum}
\end{table}

This is the main result of this work. Before we move to the next section, let us shortly discuss the above classification. The arbitrariness of the functions $g_i(X)$ makes this classification broad enough, as far as the restrictions are concerned. By choosing specific classes (and thus symmetries) and playing with the form of the function $g_i$, we can map  modifications of GR to the Horndeski theory and see if they are invariant or not under the action of Noether point symmetries. In this perspective, the Noether Symmetry Approach is a selection criterion discriminating among integrable models. As discussed in \cite{epjc},  the Noether symmetries   select  ``physical" models in the sense that the related conserved quantities  result physical observables of the theory.

Moreover, from the Noether vector \eqref{eq22}, we can define the Lagrange system
\begin{equation}\label{eq23}
\frac{dt}{\xi_1 t + \xi_2}  = \frac{da}{\eta _a(a)} = \frac{d\phi}{\xi_1 \phi + \phi_1}\,.
\end{equation}
Without loss of generality, we can set $\xi_2 = 0$. As we already mentioned before, for $g(\phi) \neq  0$ the $\eta_a$ coefficient becomes $\eta_a(a) = \alpha _1 a$, while for $g(\phi) = 0$, it is $\eta_a(a) = \frac{1}{3}(\alpha_1+2\xi_1)a$. By solving the system \eqref{eq23} for each case, we get the zero-order invariants which are solutions of the system of the E-L equations
\begin{gather}
a(t) = \alpha_0  t  ^{\alpha_1/\xi_1}\,,\,\,\phi (t) = \phi_0 t - \frac{\phi_1}{\xi_1} \, \text{for} \,\,g(\phi) \neq  0 \,, \nonumber \\
a(t) = \alpha_0   t ^{(\alpha_1+2\xi_1)/3\xi_1}\,,\,\,\phi (t) = \phi_0 t - \frac{\phi_1}{\xi_1} \, \text{for} \,\,g(\phi) = 0 \,\,. \nonumber 
\end{gather}
There are two E-L equations, one for $a$ and one for $\phi$, but we also have the constraint equation. By plugging these solutions in the E-L equations we can get constraints for the arbitrary functions $g_i(X)$ in the table \ref{tab:sum}.

\section{From Horndeski to specific  modified theories of gravity}

By choosing specific forms of the arbitrary functions $g_2(X),\,g_4(X)$ and $g_5(X)$, as well as by fixing the constants $\xi_1,\,\phi_1,\,\alpha_1$ and $c_i$, we can recast the Horndeski Lagrangian, to  Lagrangians coming from modified theories. For  each theory, if Noether symmetries exist, we can find out exact cosmological solutions.  In what follows, we match theories that show symmetries (the different classes are presented in  Table I), with some extended  theories of gravity (Brans-Dicke, $f(R)$, etc). For these theories, cosmological solutions exist and we  present them. In principle, the approach consists in finding out  the conserved quantities for  each case (if they exist), in reducing the dynamics of the system,  and in obtaining  exact solutions. 


\subsection{Brans-Dicke gravity}

Let us start with the simplest, and one of the first considered  modification of gravity, the Brans-Dicke theory. The action is  \cite{Brans:1961sx}
\begin{equation}
\mathcal{S} \sim \int d^4x \sqrt{-g} \left[ \phi R - \frac{\omega}{\phi}\nabla _{\mu}\phi \nabla^{\mu}\phi - V(\phi)\right]+\mathcal{S}_m\,
\end{equation}
where $\omega$ is the Brans-Dicke parameter, i.e. the coupling constant between the scalar field and the metric. In this theory,  the Newton constant, $G$, is not constant, but it varies according to the evolution of a scalar field  $\phi \sim 1/G$. The reasons for this choice are several. In particular, Brans and Dicke considered a theory which is in more agreement with Mach's principle, compared to GR, assuming that the gravitational coupling can depend on space and time.
In cosmology, the point-like, canonical Lagrangian takes the form
\begin{equation}\label{BD}
\mathcal{L} = - 6 a \phi \dot{a}^2 - 6 a^2 \dot{a} \dot{\phi} -\omega\frac{a^3}{\phi}\dot{\phi}^2 \,,
\end{equation}
where we considered that the potential $V(\phi) = 0$. In order to match  this Lagrangian to the Horndeski theory, we have to set in the case {\bf E} of table \ref{tab:sum},
\begin{equation}
c_1 = c_2 = c_3 = c_4 = 0\,,\,g_4(X) = 1\,,\,g_5 (X) = 0 \,,\,\phi_1 = 0 \,,\,\xi_1 = 1, \ \text{and} \  g_2(X) = -2 \omega X\,.
\end{equation}
The fact that the two Lagrangians coincide, means that, our Lagrangian inherits also the cosmological solutions found in \cite{Brans:1961sx} and \cite{OHanlon:1972ysn}, i.e.
\begin{itemize}
\item
For $\omega \geq - \frac{3}{2}$ and $\omega \neq -\frac{4}{3}$,
\begin{equation}
a(t) = a_0 \left(\frac{t}{t_0} \right)^q\,,\quad \phi (t) = \phi _0 \left(\frac{t}{t_0} \right)^{r} \,,
\end{equation}
\item
For $\omega \geq - \frac{3}{2}$ and $\omega = -\frac{4}{3}$,
\begin{equation}
a(t) = a_0 \left(\frac{t}{t_0} \right)^{\frac{2}{3}}\,,\quad \phi (t) = \phi _0 \left(\frac{t}{t_0} \right)^{-1} \,,
\end{equation}
\end{itemize}
where $a_0\,,\,\,\phi_0 $ are constants and $q = \frac{1}{3}(1-r)$, $r = \frac{1}{4+3 \omega}\left(1 \pm \sqrt{3 (3+2 \omega)} \right).$ From our point of view, 
this means that the equations of motion of Brans-Dicke theory remain invariant under the point transformations described by the Noether vector
\begin{equation}
\X = (t + \xi_2)\partial_t + \phi \partial_{\phi} \,.
\end{equation}
In addition, there is an integral of motion, which is given by
\begin{equation}
I = f_1 + a^2 \left(6 \phi\dot{a} + 2 \omega a \dot{\phi}\right)\,.
\end{equation}

\subsubsection*{String motivated gravity}
Let us now consider a string-motivated Lagrangian of the form \cite{paliathanasis,ritis}
\begin{equation}\label{stringaction}
\mathcal{S} \sim \int d^4 x \sqrt{-g} e^{-2 \phi} \left[R + 4\nabla_{\mu}\phi \nabla^{\mu} \phi - V(\phi)\right]\,.
\end{equation}
It turns out that this theory is actually a Brans-Dicke-like  theory for  specific forms of the coupling, the self-interaction potential, and a redefinition of the scalar field acting as the string-dilaton field. It interesting to include also this model in the discussion of the Horndeski theory and search for  its Noether symmetries since it has been extensively studied in  literature for several physical implications\footnote{Starting from a D-dimensional theory, e.g. the so called Polyakov action, after  compactification, we remain with only four macroscopic dimensions  ending up with the action \eqref{stringaction}. This is a simplification that allows us to study the dynamics of the degrees of freedom associated to  the four macroscopic dimensions. For details, see \cite{Easther:1995ba,Lovelace:1986kr,Fradkin:1984pq,Callan:1985ia}.}.

Assuming a FRW cosmology \eqref{eq5}, the above Lagrangian becomes
\begin{equation}
\label{string1}
\mathcal{L} =  e^{-2 \phi } \left[12 a^2 \dot{a} \dot{\phi} - 6 a \dot{a} ^2 - a^3 \left(4\dot{\phi} ^2 + V(\phi )\right)\right]\,.
\end{equation}
Besides, the Horndeski Lagrangian, with the Noether symmetry
\begin{equation}
\X= \xi_2 \partial _t + \frac{2}{3}\phi _1 \partial _a + \phi _1 \partial _{\phi}\,,
\end{equation}
i.e. $\xi _1 = 0$ and $\alpha _1 = \frac{2}{3}\phi _1 \neq 0$, becomes\footnote{We set $\alpha _1 = \frac{2}{3}\phi _1$ in order to recover the dilaton coupling from \eqref{stringaction}.}, after adopting  the symmetry class {\bf B} from Table \ref{tab:sum}, 
\begin{eqnarray}
\label{string2}
\mathcal{L} &=& a^3 e^{-2 \phi } g_2(X)  + \frac{c_2 }{3} a^3 e^{-2 \phi } \dot{\phi}^4 -c_3 a^3 e^{-2 \phi } \dot{\phi}^2  + 3 a e^{-2 \phi } \left(2 g_4'(X)-g_5(X)\right) \dot{a}^2 \dot{\phi}^2- \nonumber \\
&&- \frac{1}{2} e^{-2 \phi }  g_5'(X) \dot{a}^3 \dot{\phi}^3 - 6 a e^{-2 \phi } g_4(X) \dot{a}^2 + c_2 a^2 e^{-2 \phi } \dot{a} \dot{\phi}^3 + 12 a^2 e^{-2 \phi } g_4(X) \dot{a} \dot{\phi}\,.
\end{eqnarray}
The two actions \eqref{string1} and \eqref{string2} become  the same, if we identify
\begin{equation}
g_2(X) = - V(\phi) = V_0\,,\,\, c_1 = c_2 = c_4 = 0 \,,\,\,c_3 = 4 \,,\,\, g_4(X) = 1\,,\,\,g_5(X) = 0 \,.
\end{equation}
In this way, the Horndeski functions take the following form,
\begin{equation}
G_2(\phi,X) = V_0 e^{-2\phi} \,,\,\,G_3(\phi,X) = - 8 \phi_1 e^{-2\phi}\,,\,\,G_4(\phi,X) = e^{-2\phi}\,,\,\, G_5(\phi,X) = 0\,.
\end{equation}
As we see, the form of $V(\phi)$ is not arbitrary, and specifically, it is the constant $V_0$. Solutions in 4 dimensions are discussed in \cite{ritis,gionti,paliathanasis}. Solutions  in D dimensions are discussed in \cite{Easther:1995ba}.

\subsection{$f(R)$ gravity}
Another class of modified theories is the $f(R)$  gravity. If one replaces the Ricci scalar in the Einstein-Hilbert action, with an arbitrary function $f$ of the Ricci scalar, the family of $f(R)$ theories arise. In some sense, this is the most straightforward generalization of GR.  The arbitrariness of the function $f$ allows, in specific cases, to explain lingering problems in cosmology and astrophysics, such as the accelerated expansion, the structure formation, the inflation, etc, without including exotic forms of matter/energy in the stress-energy tensor. For the interested reader, there is a large amount of literature on this topic. For reviews  see \cite{Capozziello:2011et,Sotiriou:2008rp,DeFelice:2010aj,odintsovreport}.

As already shown in \cite{Capozziello:2011et} and references therein, by setting $\phi \equiv f'(R)\Rightarrow R = \mathcal{R}(\phi)$ and $V(\phi) = \phi \mathcal{R}(\phi) - f(\mathcal{R}(\phi))$ we obtain the following equivalence,
\begin{equation}\label{f(R)}
\mathcal {S} \sim \int d^4x \sqrt{-g} f(R) \Leftrightarrow \mathcal {S} \sim\int d^4x \sqrt{-g} \left(\phi  R - V(\phi) \right)\,.
\end{equation}
 This scalar-tensor form of $f(R)$ theories is similar to the Brans-Dicke theory, without the kinetic term, i.e. with $\omega = 0$ and with an arbitrary potential $V(\phi)$ (see \cite{OHanlon:1972sdp}). The point like Lagrangian of this action is given by
\begin{equation}\label{f(R)lag}
\mathcal{L} = - 6 a \phi \dot{a}^2 - 6 a^2 \dot{a}\dot{\phi}- a^3 V(\phi)\,,
\end{equation}
which means that in order to match it with the Horndeski Lagrangian \eqref{eq9} we have to set
\begin{equation}
G_2(\phi,X)  = - V(\phi)\,,\,\,g(\phi) = 0\,,\,\,h(\phi) = \text{const.}\,,\,\, G_4(\phi,X) = \phi\,\,\text{and}\,\,G_5(\phi,X) = 0\,.
\end{equation}
By comparing with the different classes of symmetries from the table \ref{tab:sum}, we can see that $f(R)$ can be recovered only from the {\bf C, E, G} or {\bf H} class. For example, in the {\bf E}-class we can set
\begin{equation}
\xi_1 = 1\,,\,\,\phi_1 = 0\,,\,\, g_2(X) = V_0\,,\,\, c_1 = c_2 = c_3 = c_4 = 0\,,\,\,g_4(X) = 1\,\,\text{and}\,\,g_5(X) = 0 \,,
\end{equation}
with $V_0$ an arbitrary constant and get that $V(\phi) = V_0/\phi$. This potential corresponds to the $f(R) = \sqrt{R}$ model. In fact, if we force the coupling of the scalar field with curvature to be of the form $\phi R$, we always end up with this potential and thus only with $f(R) = R^{1/2}$. However, we know from the literature \cite{cimento, Vakili:2008ea,Paliathanasis:2011jq,defelice}, that $f(R)$ accepts more Noether symmetries. Specifically, the power law model $f(R) = R^n$ accepts the Noether vector
\begin{equation}
\mathcal{X} = 2 t \partial _t + \frac{a}{3}\left(4n - 2 \right) \partial_a - 4R\partial _R\,.
\end{equation}
In order for this to be the same with the vector \eqref{eq22} we have to set $\xi_1 = 2\,,\,\, \xi_2 = 0$ and $\eta_a = a(4n-2)/3$ or better $a_1 = (4n-2)/3$ in the {\bf C} class of symmetries and $a_1 = 4n-6$ in the {\bf G} class. As an example, let us check the $n = 3/2$ case, which accepts a symmetry \cite{Paliathanasis:2011jq}. For $n=3/2$ it is $a_1 = 4/3$ (if we consider the {\bf C} class of symmetries) and thus the Horndeski functions should be
\begin{equation}
G_4(\phi,X) = (2\phi)^{-1}\,\,\text{and}\,\, G_2(\phi,X) = \frac{V_0}{8\phi^3}\,,
\end{equation}
where for simplicity we set $\phi_1 = 0$. Now the Lagrangian density looks like $\mathcal{L} \sim R/(2\phi)- V_0/(8\phi^3)$, but if we redefine the scalar field as $\psi = 1/(2\phi)$ it becomes
\begin{equation}
\mathcal{S} \sim \int d^4x \sqrt{g}\left( \psi R - V_0 \psi^3\right)\,. 
\end{equation}
In this way we can recover the power-law $f(R)$ models that admit symmetries.

As discussed in \cite{arturo} for spherical symmetry, the power $n$ is related to the conserved quantities that have physical meaning \cite{vesna,sergio}. It is straightforward to solve the Euler-Lagrange equations produced by \eqref{f(R)lag} to get
\begin{equation}
a(t) = a_0 \left(\frac{t}{t_0}\right)^m\,,\quad \phi(t) = \pm i \sqrt{\frac{V_0}{48 m^2 - 24 m}}t\,.
\end{equation}
In order for the scalar field solutions to be real, we have two branches: 1) $V_0<0$ and $0<m<1/2$ and 2) $V_0>0$ and $m<0 \, \text{or}\, m>1/2$. There exist also exponential solutions for the scale factor, which lead to constant scalar field.

\subsection{Cubic Galileon model}
The galileon theories have also been proposed as an natural explanation of the accelerated expansion of the Universe, without the need of dark energy and, as such, a lot of progress has been made in the last few years in this direction. The name comes from the fact that, in galileon gravity theories, the  action is invariant under the shift symmetry in flat spacetime, $\partial _{\alpha} \phi \rightarrow \partial _{\alpha} \phi + \upsilon _{\alpha}$. They pass the Solar-System tests  \cite{Vainshtein} and applications of MOND have been studied in this context \cite{Babichev:2011kq}. Inflationary and self-accelerating solutions have been also been considered  \cite{Kobayashi:2010cm,Deffayet:2010qz,Silva:2009km,Kobayashi:2010wa,Burrage:2010cu} and, moreover, gravitational waves have also been taken into account \cite{Chu:2012kz,deRham:2012fw}.
We will focus on the cubic galileon theory with the action, in the Einstein frame, given by
\begin{equation}\label{cubicgalileon}
\mathcal{S} \sim \int d^4x \sqrt{-\tilde{g}}\left[\tilde{R}-\frac{k_1}{2}\tilde{\nabla}_{\mu}\psi\tilde{\nabla}^{\mu}\psi - \frac{k_2}{2 M^2}\tilde{\nabla}_{\mu}\psi\tilde{\nabla}^{\mu}\psi \tilde{\square} \psi  \right] +\mathcal {S}[\chi_m,g_{\mu\nu}]\,.
\end{equation}
The spacetime metric is described by $\tilde{g}_{\mu\nu}$, $k_1,\,k_2$ are coupling parameters and $M$ is a mass scale of the galileon field, $\psi$. Matter fields, $\chi_m$, couple minimally to a physical metric (in the Jordan frame) $g_{\mu\nu} = e^{2\alpha \psi} \tilde{g}_{\mu\nu}$, with $\alpha$ the matter-galileon coupling parameter \cite{Bhattacharya:2015chc}.

Matching  the Einstein-cubic galileon and the Horndeski theory, i.e symmetry class {\bf A} in the Table \ref{tab:sum},  we have to set
\begin{equation}
 g_2(X) = k_1 X\,,\,g_4(X) = 1\,,\, g_5(X) = 0\,,\, c_1 =0 \,,\, c_2 = \frac{k_2}{M^2}\,,\,c_3 =0\,,\,c_4 = 0\,,
\end{equation}
where the Noether vector takes the form 
\begin{equation} 
\X = \xi_2 \partial_t + \phi_1 \partial_{\phi}\,,
\end{equation} 
and the integral of motion becomes
\begin{equation}
I = f_1 - \phi _1 a^2  \dot{\psi} \left(k_1 a -\frac{3 k_2}{M^2} \dot{a} \dot{\psi}\right)\,,
\end{equation}
since the point-like cosmological Lagrangian coming from \eqref{cubicgalileon} is
\begin{equation}
\mathcal{L} = - 6 a \dot{a}^2 + \frac{k_1}{2} a^3 \dot{\psi} -\frac{k_2}{M^2}  a^2 \dot{a}\dot{\psi}\,.
\end{equation}
This model is very well studied in the literature and there have been found both cosmological as well as spherically symmetric solutions \cite{Babichev:2012re,Bhattacharya:2015chc}. For example, if one considers the linear {\it ansatz} 
\begin{equation}
\phi (t) = \phi_0 + \phi _1 t\,,
\end{equation} 
where $\phi_0$ and $\phi_1$ are constants, for the scalar field, they get that $H = k_2 M^2/(3 k_3 \phi_1)$, which is an expanding solution as long as $k_2k_3\phi_1 > 0$.


\subsection{Non-minimal kinetic coupling}
An interesting subclass of Horndeski theory is represented by scalar-tensor
models where the scalar kinetic term has non-minimal coupling to curvature.
Theories with the non-minimal kinetic coupling lead to a rich variety of solutions
for  different cosmological epochs, particularly for late time acceleration,
as shown in \cite{Sus:2009, SarSus:2010, Sus:2012, StaSusVol:2016, MatSus:2018, GubLin:2011,schmidt1}.

The action of the theory of gravity with non-minimal kinetic coupling reads
\begin{equation}\label{nonmincouplaction}
S=\int d^4x\sqrt{-g}\left\{ \frac{R}{16\pi} -\frac12\big[g^{\mu\nu} + \eta G^{\mu\nu} \big] \nabla_\mu\phi \nabla_\nu\phi -V(\phi)\right\},
\end{equation}
where $\eta$ is a coupling parameter with the dimension of inverse mass-squared.
Comparing this with the Horndeski action (\ref{eq1}), we find
\begin{equation}
G_2(\phi,X) = X-V(\phi)\,,\,\,G_3(\phi,X) = 0\,,\,\,G_4(\phi,X) = \frac{1}{16\pi}\,,\,\,G_5(\phi,X) = \frac{1}{2}\eta\phi.
\end{equation}
Since we assume that $G_3(\phi,X)=0$, then from Eq.\eqref{G3XX} we get $g(\phi)=0$ and $h(\phi)=0$. In addition, the coupling to the Einstein tensor is derived by integrating out a total derivative. The theory \eqref{nonmincouplaction} possesses the Noether symmetry iff $V(\phi)\sim\Lambda=const$, and the configuration providing the Noether symmetry belongs to the symmetry-class {\bf J} in Table \ref{tab:sum}, where
\begin{equation}
c_1 =0 =c_2\,,\,\,c_3=\frac{1}{2}\eta\,\,\,g_2(X) = X -2\Lambda\,,\,\, g_4(X)=\frac{1}{16\pi}\,,\,\, g_5(X)=0.
\end{equation}
Now, the Lagrangian \eqref{eq9} takes the form
\begin{equation}
\mathcal{L} = a^3 ({\textstyle\frac12}\dot\phi^2-2\Lambda) -\frac{3a\dot{a}^2}{8\pi} -\frac{3}{2}\eta a {\dot a}^2 {\dot{\phi}}^2.
\end{equation}
After solving the Euler-Lagrange equations for the above Lagrangian, we get, e.g. for $\Lambda >0$ and $\eta >0$, 
\begin{gather}
a(t) = H_{\Lambda}t\,,\,\phi(t) = \phi_0 = \text{const.} \\
a(t) = \frac{t}{\sqrt{3\eta}}\,,\,\, \phi(t) = \sqrt{\frac{3\eta H^2_{\Lambda}-1}{16 \pi \eta}}t\,,
\end{gather}
where $H_{\Lambda} \geq 1/\sqrt{3\eta}$. For different combinations of   $\Lambda$ and $\eta$ signs, as well as for a  discussion on  solutions (e.g. with $\Lambda = 0$), see \cite{SarSus:2010} and references therein.

\section{Discussion and Conclusions}
The Horndeski gravity  is the most general scalar-tensor theory   giving rise to  second order field equations. In principle, any theory of  gravity containing scalar-tensor  terms can be mapped onto the action \eqref{eq1}.  In this paper,  we proposed a systematic classification of  scalar-tensor models coming from the Horndeski theory  which are  invariant under infinitesimal point transformations. Specifically, using the so-called Noether Symmetry Approach, we were able to find  theories that possess symmetries and thus, integrals of motion. When symmetries exist, the related dynamical systems are reducible and integrable. In other words, the presence of  symmetries fixes the functional form of the theory, gives conserved quantities and allows to find out exact solutions. 

In Table \ref{tab:sum}, we reported all the  FRW cosmologies, derived from the Horndeski gravity, by Noether symmetries. As it appears evident, the existence of Noether symmetry fixes the classes of models and their mathematical and physical properties. 

The paradigm is twofold: $i)$  couplings and  scalar-field potentials of a given theory can be derived from the general Horndeski action \eqref{eq1}; $ii)$ the invariance under  point infinitesimal   transformations gives rise to the Noether symmetries and then allows to exactly integrate the system.
Furthermore,  the most popular alternative gravities come out from this approach and can be worked out under the standard of Noether symmetries. In particular, we considered Brans-Dicke gravity, $f(R)$ gravity, galileon gravity, string motivated gravity and non-minimal derivative coupling gravity. They are  five specific models of theories  belonging  to the  four classes of the Noether symmetry: {\bf A}, {\bf B}, {\bf E}, and {\bf J}.  In principle, all symmetry classes can be discussed  under the present standard.

An important remark  is necessary at this point. In the last two years, significant   progress has been done in  gravitational wave astronomy. Specifically, the observation of black hole-black hole mergers, as well as the binary neutron star merger GW170817 \cite{multi}, have provided the possibility to test GR in the strong field regime. The last  observed event (binary neutron stars), together with its electromagnetic counterpart, started the so-called  the {\it multi-messenger astrophysics}  setting severe constraints on the propagation of tensor modes. Since the Horndeski theory shows, besides the standard   $+$ and $\times$ polarization modes of GR, an extra mode excited by a massive scalar field \cite{Hou:2017bqj}, it means that the theory can be  severely constrained by the mass of the graviton \cite{Kreisch:2017uet,Gong:2017kim}. Besides, the motion of stars as well as the energy radiated away as  gravitational radiation  are different if compared to GR:  this means that more constraints can be  obtained  and several Horndeski models can be ruled out by the observations \cite{Hou:2017cjy}. 
In particular,  some  models (such as the non-minimal derivative coupling) are presently excluded by gravitational wave observations and then $G_4$ and $G_5$ functions  are strictly constrained. However, also considering observational limitations, our approach goes beyond because it is aimed to classify the general  Horndeski action. 

As we already mentioned, the purpose of this article is to classify all the possible models originating from the general Horndeski action \eqref{eq1}, that present Noether symmetries. Clearly the zero-order invariants, derived from symmetries, can be used to construct general  exact solutions. For example,  in Refs.  \cite{Fre:2013vza} and \cite{Kamenshchik:2013dga}, cosmology coming from scalar-tensor theories of gravity have been discussed in detail deriving exact solutions from zero-order invariants.
In particular, in Tables I and II of Ref.\cite{Kamenshchik:2013dga}, the specific forms of gravitational coupling and self-interaction potential are given  allowing to achieve the general exact solutions for the scalar-tensor dynamics related to their action (1). Such an action, can be derived, from our approach, specifying, for example, the form of function $G_2$. In other words, our Table I can be compared to Tables I and II in [84] deriving the same results. Similar considerations hold for \cite{Fre:2013vza}.   In a future work, we will study physically interesting theories for each class of models and use the related zero-order invariants, i.e.  the  conserved quantities, to reduce dynamics and find out exact solutions. 
Moreover, following the approach reported in   \cite{leandros}, we will use cosmological observations in order to constrain the parameters of Noether symmetries in Horndeski gravity.

\section*{Acknowledgments}
S.C. and K.F.D. acknowledge the support of INFN (iniziative specifiche QGSKY and TEONGRAV). S.V.S. is supported by the RSF grant 16-12-10401. K.F.D. thanks also Sebastian Bahamonde for interesting discussions. The authors would like to thank the anonymous referee for her/his helpful comments. This  paper is based upon work from COST action CA15117  (CANTATA), supported by COST (European Cooperation in Science and Technology).

\end{document}